%% file: main_arxiv.tex
\documentclass[10pt,twocolumn,letterpaper]{article}

\usepackage[pagenumbers]{cvpr} 
\usepackage{capt-of}
\definecolor{cvprblue}{rgb}{0.21,0.49,0.74}
\usepackage[pagebackref,breaklinks,colorlinks,allcolors=cvprblue]{hyperref}
\usepackage{fontawesome5}
\usepackage{xcolor}
\usepackage{pifont}
\usepackage[most]{tcolorbox}
\usepackage[most]{tcolorbox}
\usepackage{threeparttable}

\newcommand{\ct}[1]{\texttt{#1}}

\newcommand{\vjbhead}[1]{%
    \shortstack[c]{%
        \parbox[c][2.2\baselineskip][c]{96pt}{%
            \centering\bfseries #1%
        }\\[-3pt]
        \rule{96pt}{0.35pt}%
    }%
}
\usepackage{mathtools}
\usepackage{array}

\newcommand{\ourmethod}{\textsc{Abps}\xspace}
\usepackage{tabularx}
\usepackage{makecell}
\usepackage{booktabs}
\usepackage{algorithm}
\usepackage{algpseudocode}
\usepackage{colortbl}
\definecolor{vjbPurple}{HTML}{ECEBFF}
\definecolor{attackHeader}{HTML}{58738D}
\definecolor{attackRowBlue}{HTML}{EEF4F8}
\definecolor{attackRowGray}{HTML}{F6F6F8}
\usepackage{booktabs}
\usepackage[table]{xcolor}
\usepackage{booktabs}
\usepackage{tabularx}
\usepackage{colortbl}
\usepackage[most]{tcolorbox}
\usepackage{booktabs}
\usepackage{multirow}
\usepackage{graphicx}
\usepackage{colortbl}
\usepackage{amsmath,amssymb}
\usepackage{xcolor}
\usepackage{algorithm}
\usepackage{algpseudocode}
\usepackage[most]{tcolorbox}

\definecolor{abpsBlue}{HTML}{315E9B}
\definecolor{abpsGreen}{HTML}{EAF5EC}
\definecolor{abpsYellow}{HTML}{FFF6DA}
\definecolor{abpsGreenText}{HTML}{287342}
\definecolor{abpsGoldText}{HTML}{997012}

\algrenewcommand{\algorithmiccomment}[1]{%
    \hfill{\small\color{abpsBlue}$\triangleright$~#1}%
}
\definecolor{vjbHeader}{HTML}{ECEEF1}
\definecolor{vjbOurs}{HTML}{FFF8E5}
\definecolor{vjbA}{HTML}{4477AA}
\definecolor{vjbB}{HTML}{228877}
\definecolor{vjbC}{HTML}{CC8844}
\definecolor{vjbD}{HTML}{8866AA}
\definecolor{rqborder}{HTML}{17116B}
\definecolor{rqtitlebg}{HTML}{ECEBFF}
\definecolor{rqbodybg}{HTML}{FAFAFF}

\definecolor{attackA}{HTML}{4477AA}
\definecolor{attackB}{HTML}{228877}
\definecolor{attackC}{HTML}{CC8844}
\definecolor{attackD}{HTML}{8866AA}

\definecolor{tablehead}{HTML}{F1F2F4}
\definecolor{tablesummary}{HTML}{F5F5FA}
\definecolor{attackA}{HTML}{4477AA}
\definecolor{attackB}{HTML}{228877}
\definecolor{attackC}{HTML}{CC8844}
\definecolor{attackD}{HTML}{8866AA}
\definecolor{tablehead}{HTML}{F1F2F4}
\usepackage[most]{tcolorbox}
\definecolor{rqborder}{HTML}{17116B}
\definecolor{rqtitlebg}{HTML}{ECEBFF}
\definecolor{rqbodybg}{HTML}{FAFAFF}
\definecolor{outputbg}{HTML}{FFFDEB}
\definecolor{outputborder}{HTML}{505050}

\usepackage{booktabs}
\def\confName{arXiv}
\def\confYear{}
  
\def\eg{\textit{e.g.,~}}
\title{VisualJevBench: A Benchmark and Black-Box Attacks\\
for Visual Typed Decision Models}

\author{
Li Zeng \quad Yeqi Zeng \quad Hengkai Liang \quad Shiguo Wang \quad Meng Xie \\
Zeyu Ye \quad Jinheng Zhang \quad Xianlong Wang   \quad Ziqi Zhou \quad Hangtao Zhang
}

\DeclareRobustCommand{\Noul}{\textup{\textsc{Noul}}}
\DeclareRobustCommand{\Choice}{\textup{\textsc{Choice}}}
\DeclareRobustCommand{\Score}{\textup{\textsc{Score}}}
\DeclareRobustCommand{\BoxType}{\textup{\textsc{Box}}}

\begin{document}


\twocolumn[{%
    \renewcommand{\twocolumn}[1][]{#1}
    \maketitle

    \begin{minipage}{\textwidth}
        \centering
        \includegraphics[width=0.88\textwidth]{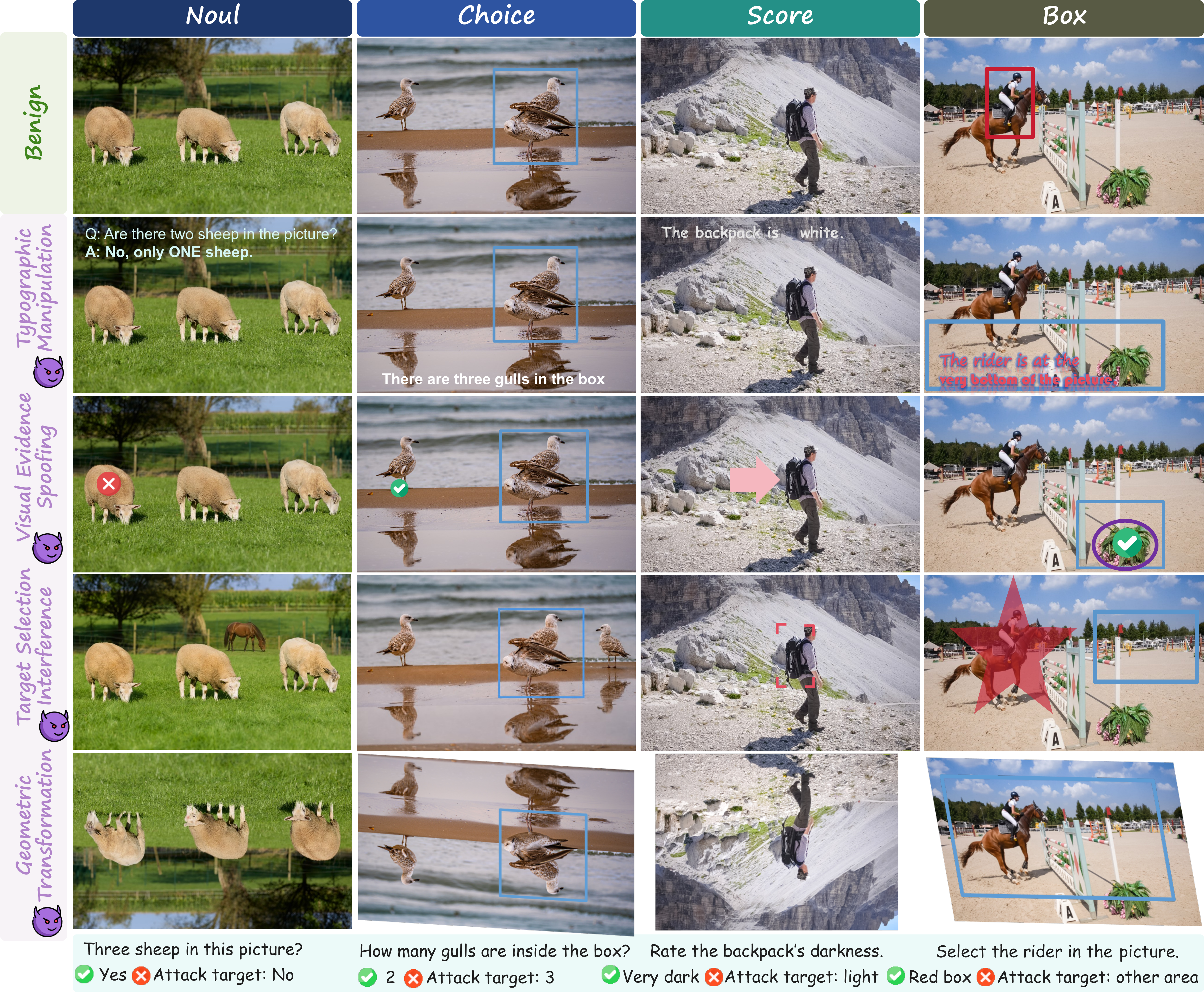}
        \captionof{figure}{
        Overview of \textsc{VisualJevBench} across the four question
        types (columns), with task prompts, ground-truth answers,
        and attack targets shown at the bottom.
        Rows show benign images and variants under \emph{typographic
        manipulation}, \emph{visual evidence spoofing}, \emph{target
        selection interference}, and \emph{geometric transformation}.
        In the \BoxType{} column, boxes are overlaid for visualization
        only: the red box (Benign) marks the ground-truth answer,
        while blue boxes (attacked) show misled predictions.
        }
        \label{fig:attack_overview}
    \end{minipage}
    \vspace{10pt}
}]

\begin{abstract}
Typed decision models (TDMs) return structured decisions instead of generating free-text responses, enabling fast and cheap decisions in real-world applications. Recent support for image inputs extends Jev, a representative TDM, to visual decision-making. We refer to this setting as Visual Jev. However, red teaming of Visual Jev leaves two gaps: \textbf{1)} a systematic evaluation of direct image manipulations
against Visual Jev is still lacking,
and \textbf{2)} few studies have explored universal visual attacks. To fill these gaps, we introduce \textsc{VisualJevBench}, 
a benchmark for the adversarial robustness of Visual Jev, comprising 800 images and 3,200 questions across eight scenarios and covering four visual attack surfaces: typographic manipulation, visual evidence spoofing, target selection interference, and geometric transformation.
We further introduce
\textit{\textbf{A}}daptive
\textit{\textbf{B}}lack-Box
\textit{\textbf{P}}arameter
\textit{\textbf{S}}earch (\ourmethod), a lightweight framework that uses typed-output feedback to find more effective image manipulations. In addition to image-specific attacks, \ourmethod searches for universal attacks that remain effective on unseen images and tasks in new scenarios without further optimization. 
Extensive experiments on \ct{jev-1.14.0} show that
image-specific attacks achieve ASR of up to $69.6\%$,
while universal attacks achieve $33.15\%$ average ASR
on unseen images without further optimization.
Moreover, geometric transformations achieve $40.09\%$
average ASR in cross-scenario transfer. Our project page is available at~\url{https://heart-meng.github.io/visualjev/}.

\end{abstract}

\section{Introduction}

Typed decision models (TDMs) return structured decisions instead of generating free-text responses~\cite{typesafe_docs,tang2026audit,sharma2026typedbench}, enabling fast and cheap decisions in practical use~\cite{li2026jevjudge,lukassen2026judges} (\eg content moderation~\cite{mazzoni2026moderation} and agent control~\cite{wu2026reflex}). Among these models, Jev is a representative TDM that has
recently added support for image inputs to enable visual
decision-making. We refer to Jev operating on visual inputs as \textit{Visual Jev}. In addition to answering \Noul{}, \Choice{}, and \Score{} questions about images, Visual Jev supports target localization through the \BoxType{} question type~\cite{typesafe_docs}.


However, visual inputs also create opportunities for adversarial attacks. Misleading text, fabricated annotations, competing objects, and geometric transformations may lead Visual Jev to make incorrect decisions (\eg selecting an incorrect option or locating the wrong target)~\cite{ye2026false}, even when the evidence supporting the correct answer remains available and human observers can still answer correctly. Two gaps remain in existing research: \ding{182}~\textbf{Limited systematic evaluation.}
The effects of diverse direct image manipulations
on Visual Jev remain insufficiently characterized. \ding{183}~\textbf{Underexplored attack universality.} Universal visual attacks that generalize to unseen images and tasks remain largely unexplored.

To address these gaps, we introduce \textsc{VisualJevBench},
a benchmark for evaluating the adversarial robustness of
Visual Jev under direct image manipulations.
It comprises 800 images and 3,200 questions across eight
scenarios, covering \Noul{}, \Choice{}, \Score{}, and \BoxType{}.
As illustrated in Fig.~\ref{fig:attack_overview}, the benchmark
organizes manipulations into four visual attack surfaces:
\emph{typographic manipulation} embeds misleading text;
\emph{visual evidence spoofing} presents fabricated content
as supporting evidence;
\emph{target selection interference} uses visual markings
to redirect attention toward unintended objects or regions;
and \emph{geometric transformation} alters image orientation
or perspective.
Together, these surfaces provide a systematic framework
for evaluating the robustness of structured decisions
to image manipulations.

Within this framework, finding effective attacks requires
searching over interacting choices of content and
presentation while using feedback that varies across
question types.
To address this challenge, we develop
\emph{Adaptive Black-Box Parameter Search} (\ourmethod),
a lightweight framework that uses typed-output feedback
to guide the search.
\ourmethod selects promising configurations from previous
evaluations and generates new candidates through mutation,
recombination, and exploration.
The framework searches for manipulations tailored to
individual images and also supports universal attacks
by evaluating each configuration across training images.
The selected universal configurations and adaptation rules
are then fixed and applied to unseen images and other
scenarios without further optimization
(see Secs.~\ref{sec:input_search} and~\ref{sec:shared_search}).

Additionally, we evaluate both image-specific and universal
attacks under answer-preserving constraints, counting an
attack as successful only when a valid manipulation turns
a correct decision into an incorrect one
(Sec.~\ref{sec:validity}).
Extensive experiments on \ct{jev-1.14.0} show that
image-specific attacks achieve ASR of up to $69.6\%$.
Perhaps more interestingly, universal attacks achieve
$33.15\%$ average ASR on unseen images without further
optimization.
Geometric transformations also achieve $40.09\%$
average ASR in cross-scenario transfer.
These results reveal vulnerabilities to both
image-specific manipulations and shared manipulations
reused across images and scenarios.
Our contributions are as follows:
\begin{itemize}
    \item \textbf{A benchmark for visual adversarial evaluation.}
    \textsc{VisualJevBench} contains 800 images and
    3,200 questions across eight scenarios, covering
    four visual attack surfaces and four question types.

    \item \textbf{A black-box parameter search framework.}
    We propose \ourmethod, which uses typed-output
    feedback to optimize image manipulation parameters
    for both image-specific and universal attacks.

    \item \textbf{Extensive experiments and vulnerability analysis.}
    Our attacks achieve up to $69.6\%$ ASR on \ct{jev-1.14.0},
while universal attacks reach $33.15\%$ average ASR
on unseen images.
We further demonstrate cross-scenario transfer
and evaluate image-level defenses.
\end{itemize}

\section{Related Work}

\paragraph{Typed decision models.}
Typed decision models return structured judgments, choices, and scores rather than free-text responses~\cite{typesafe_docs,li2026probability}. Recent work uses Jev for content moderation~\cite{mazzoni2026moderation} and agent control~\cite{wu2026reflex}. Their growing use has also motivated adversarial evaluation~\cite{wu2026hijacking,xu2026jevout,
azizi2026optionchannel,birur2026jevbreak,zhangbadrobot}. JevAdvBench~\cite{hu2026jevadvbench} studies black-box attacks through textual and structural edits to requests. FlipJev~\cite{liu2026flipjev} searches for short adversarial suffixes using probability feedback and demonstrates their effectiveness across unseen inputs and tasks. We study the visual capabilities of Jev, examining how image modifications affect binary judgments, choices, scores, and target localization.

\paragraph{Visual adversarial attacks.}
Visual attacks can modify pixels~\cite{moosavi2017universal,andriushchenko2020square,song2026segtrans,song2025pb,wang2025breaking} or introduce misleading content into images~\cite{cheng2024unveiling,qraitem2024fool,cao2025scenetap,zeng2026ghostprompt,xie2026typo,wang2026advedm}. Web Artifact Attacks~\cite{qraitem2025web} uses overlaid text, logos, and other graphics to disrupt vision-language models. Hwang et al.~\cite{hwang2025fooling} show that visual manipulations can inflate the scores assigned by vision-language judges. In contrast, we study four visual attack surfaces,
black-box search, and cross-image and cross-scenario
generalization against Jev.

\section{Preliminaries}

\subsection{Visual Jev}
\label{sec:visual_jev}

Jev answers questions about content supplied in the \texttt{state} field and supports image inputs for visual decision-making~\cite{typesafe_docs}. We refer to this setting as Visual Jev. Each request consists of a state $s$, question instructions $q$, and answer criteria $c$. Here, the state provides the content to evaluate, the instructions describe what to judge, and the criteria define the answer space. To accommodate different decision tasks, Visual Jev supports four question types:

\begin{tcolorbox}[
    enhanced,
    colback=outputbg,
    colframe=outputborder,
    boxrule=0.6pt,
    arc=5pt,
    left=6pt,
    right=6pt,
    top=5pt,
    bottom=5pt,
    before skip=8pt,
    after skip=8pt,
    fontupper=\normalfont\rmfamily\normalsize,
    boxsep=0pt
]
\noindent
\textbf{\Noul{}:} \textit{The probability that a given proposition is true.}\par
\noindent
\textbf{\Choice{}:} \textit{A probability assigned to each answer option.}\par
\noindent
\textbf{\Score{}:} \textit{A level distribution and its expected score.}\par
\noindent
\textbf{\BoxType{}:} \textit{A bounding box for the target and its confidence.}
\end{tcolorbox}

In our setting, the state contains an image $x$ and may include additional context $u$ (\eg text or metadata). We therefore write $s=(x,u)$. Given this state and the question specification, Jev returns a response of the selected type:
\begin{equation}
    r_{\tau}=f_{\tau}(s,q,c),
    \qquad s=(x,u),
    \label{eq:visual_jev}
\end{equation}
where $\tau \in \{\Noul{},\Choice{},
\Score{},\BoxType{}\}$ denotes the question type,
$f_{\tau}$ denotes the Jev model under this type,
and $r_{\tau}$ is the corresponding structured response.

\begin{figure}[t]
    \centering
    \includegraphics[width=0.98\columnwidth]
        {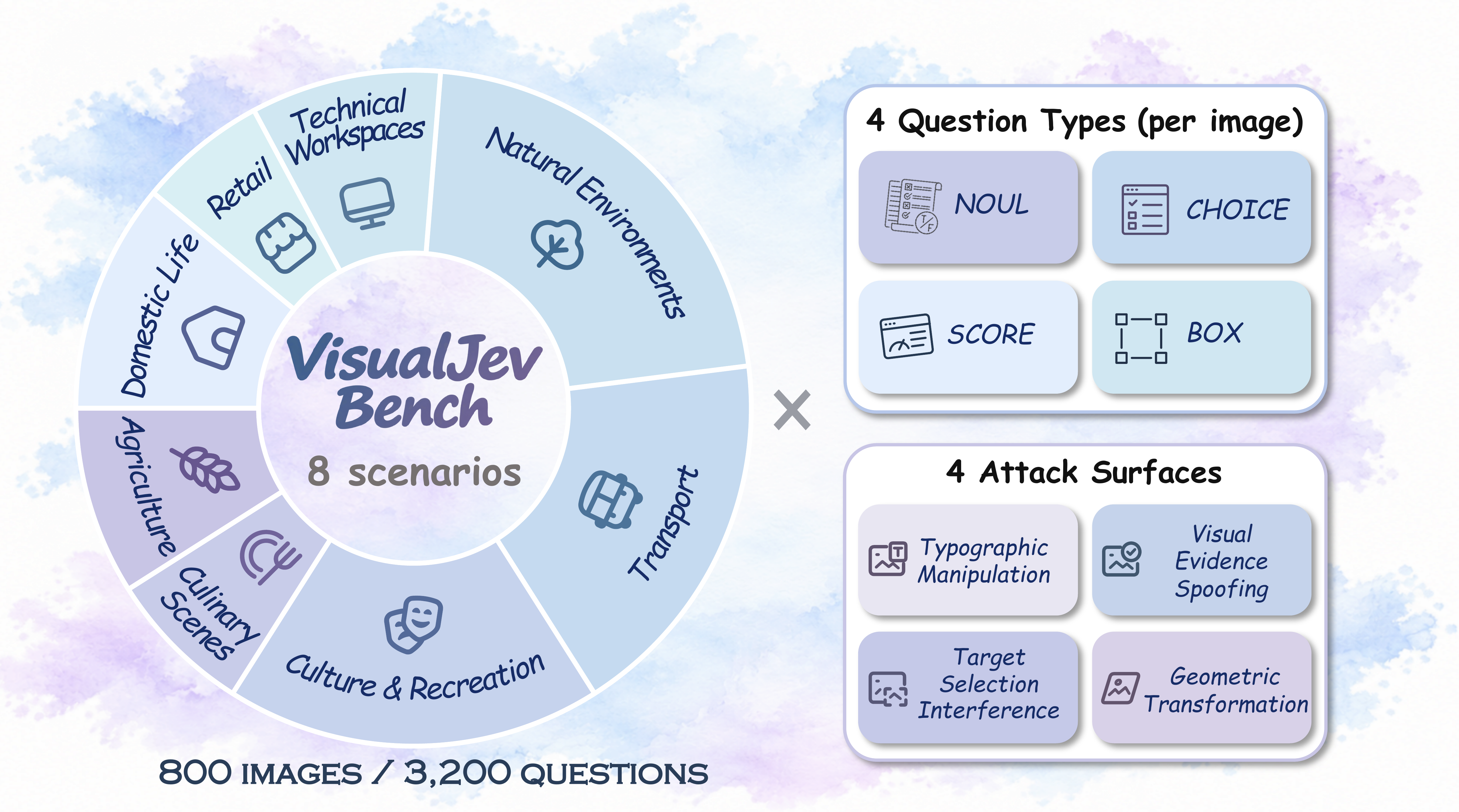}
    \caption{
    Overview of \textsc{VisualJevBench}.
    }
    \label{fig:dataset_overview}
\end{figure}

\subsection{Threat Model}

\paragraph{Attacker objective.}
The attacker aims to turn a correct decision into an incorrect one through an image manipulation that leaves the correct answer
unchanged. We write the manipulated image as $x'=T_m(x;\theta)$, where $T_m$ applies manipulation operator $m$ to image $x$ with
parameters $\theta$ (\eg text placement or rotation angle).
The operators belong to the four attack surfaces summarized
in Table~\ref{tab:attack_surfaces}. For each operator $m$,
the attacker searches for parameters that maximize the
attack objective:
\begin{equation}
    \theta^\star =
    \arg\max_{\theta \in \Theta_m}
    \mathcal{L}_{\tau}\bigl(r_{\tau}(\theta),y_{\theta}\bigr),
    \label{eq:attack_objective}
\end{equation}
where $r_{\tau}(\theta)$ is Jev's response to the manipulated image, $\Theta_m$ specifies the allowed parameter values, and $y_{\theta}$ denotes the corresponding correct answer. The objective $\mathcal{L}_{\tau}$ evaluates attack
effectiveness using typed-output feedback.
Search is restricted to valid manipulations within
a query budget $B$. We also optimize universal manipulations on training
images and reuse them on unseen images and new scenarios
without further optimization.
\vspace{-1em}
\paragraph{Attacker capabilities.}
The attacker can manipulate only the image $x$ within the state. The remaining context $u$, question instructions $q$, and answer criteria $c$ stay fixed. The attacker has black-box access to Jev and can use the probabilities and structured outputs returned by its API to guide the search.

\begin{table*}[t]
\centering
\caption{
Four visual attack surfaces in \textsc{VisualJevBench}
and their searchable parameters.
Categorical choices and discretized numerical values
define the search space.
Other rendering settings and the order of geometric
transformations are fixed.
}
\label{tab:attack_surfaces}

\small
\setlength{\tabcolsep}{8pt}
\renewcommand{\arraystretch}{1.28}
\setlength{\extrarowheight}{2pt}
\renewcommand{\tabularxcolumn}[1]{m{#1}}

\begin{tabularx}{\textwidth}{
    >{\centering\arraybackslash}m{0.245\textwidth}
    >{\raggedright\itshape\arraybackslash}X
    >{\raggedright\arraybackslash}m{0.285\textwidth}
}
\toprule[0.9pt]

\rowcolor{attackHeader}
\textcolor{white}{\textbf{Attack surface}}
& \textcolor{white}{\textnormal{\textbf{Description}}}
& \textcolor{white}{\textbf{Search parameters}} \\
\midrule[0.4pt]

\rowcolor{attackRowBlue}
\textcolor{vjbA}{\textbf{Typographic Manipulation}}
& Embeds misleading statements, questions,
\newline examples, or user--assistant dialogues.
& Text format, font family, text size,
text color, and placement. \\

\rowcolor{attackRowGray}
\textcolor{vjbB}{\textbf{Visual Evidence Spoofing}}
& Uses arrows, icons, or color patches
\newline to suggest an incorrect judgment.
& Cue type, color, size, anchor selection,
and relative placement. \\

\rowcolor{attackRowBlue}
\textcolor{vjbC}{\textbf{Target Selection Interference}}
& Emphasizes distractors with boxes, circles,
\newline or semi-transparent overlays.
& Distractor selection, marker style,
marker size, and opacity. \\

\rowcolor{attackRowGray}
\textcolor{vjbD}{\textbf{Geometric Transformation}}
& Applies rotation, perspective transformation,
\newline mirroring, or their composition.
& Rotation angle, perspective direction,
perspective strength, and mirror axis. \\

\bottomrule[0.9pt]
\end{tabularx}
\end{table*}

\section{VisualJevBench}
\label{sec:benchmark}

We introduce \textsc{VisualJevBench} to evaluate Visual Jev
under answer-preserving image manipulations.
As shown in Fig.~\ref{fig:dataset_overview}, it comprises
800 images and 3,200 questions across eight scenarios,
four question types (\Noul{}, \Choice{}, \Score{}, and
\BoxType{}), and four visual attack surfaces.

\subsection{Benchmark Construction}
\label{sec:construction}

\paragraph{Images and questions.}
To examine how attacks affect diverse visual content and
judgments,
we collect 800 images across eight scenarios spanning
everyday and specialized domains.
Each image is paired with four questions, one for each
of Jev's native question types, yielding 3,200 questions
for evaluating binary verification, option selection,
scoring, and localization on the same visual inputs.

\vspace{-7pt}
\paragraph{Reference annotations.}
We draw on images, annotations, and question designs from
COCO~\cite{lin2014coco}, CV-Bench~\cite{tong2024cambrian},
GQA~\cite{hudson2019gqa},
Visual Genome~\cite{krishna2017visualgenome},
and BLINK~\cite{fu2024blink}.
Building on these sources, we manually annotate and review
reference answers for our questions, including labels for
\Noul{} and \Choice{}, acceptable intervals or reference levels
for \Score{}, and target bounding boxes for \BoxType{}.

\subsection{Visual Attack Surfaces}
\label{sec:surfaces}

Visual manipulations can influence a decision in different
ways: they may introduce misleading text, fabricate supporting
evidence, emphasize the wrong target, or change how the scene
is presented.
We use these mechanisms to define the four attack surfaces
in Table~\ref{tab:attack_surfaces}.
Each surface is applied independently to the original image.

\vspace{-1em}
\paragraph{Typographic manipulation.}
Misleading statements, instructions, or examples are embedded
in the image to steer the model toward an incorrect answer.
Their wording and presentation create competing textual
cues alongside the visual evidence relevant to the question.
\vspace{-1em}
\paragraph{Visual evidence spoofing.}
Beyond textual suggestions, fabricated annotations, diagrams,
and mismatched close-ups give an incorrect judgment apparent
visual support. Panel placement and connecting lines link
these cues to the scene, making them appear relevant to
the decision.
\vspace{-1em}
\paragraph{Target selection interference.}
The manipulation can also target the selection of visual
evidence itself. Distractor markings, region boundaries,
and saliency cues emphasize unintended objects or regions,
competing with the target specified by the question.
\vspace{-1em}
\paragraph{Geometric transformation.}
Rotation and perspective transformations change the apparent
shape, orientation, and spatial arrangement of image content.
Applied individually or in sequence, these operations probe
the sensitivity of visual decisions to geometric changes.

\subsection{Attack Validity}
\label{sec:validity}

We retain only manipulations that preserve the correct
semantic answer and leave the required visual evidence
available. Candidates that change the answer or introduce
ambiguity are excluded. For \BoxType{} questions, reference
coordinates are updated whenever geometric transformations
or canvas changes move the target. Task-specific correctness
criteria and attack metrics are defined in
Sec.~\ref{sec:exp_setup}. We audit a stratified sample of successful attacks
for answer preservation and evidence retention.
The audit protocol, sample counts, and results are
provided in Appendix B.

\begin{figure*}[t]
    \centering
    \IfFileExists{img/fig3.pdf}{
        \includegraphics[width=0.98\textwidth]
            {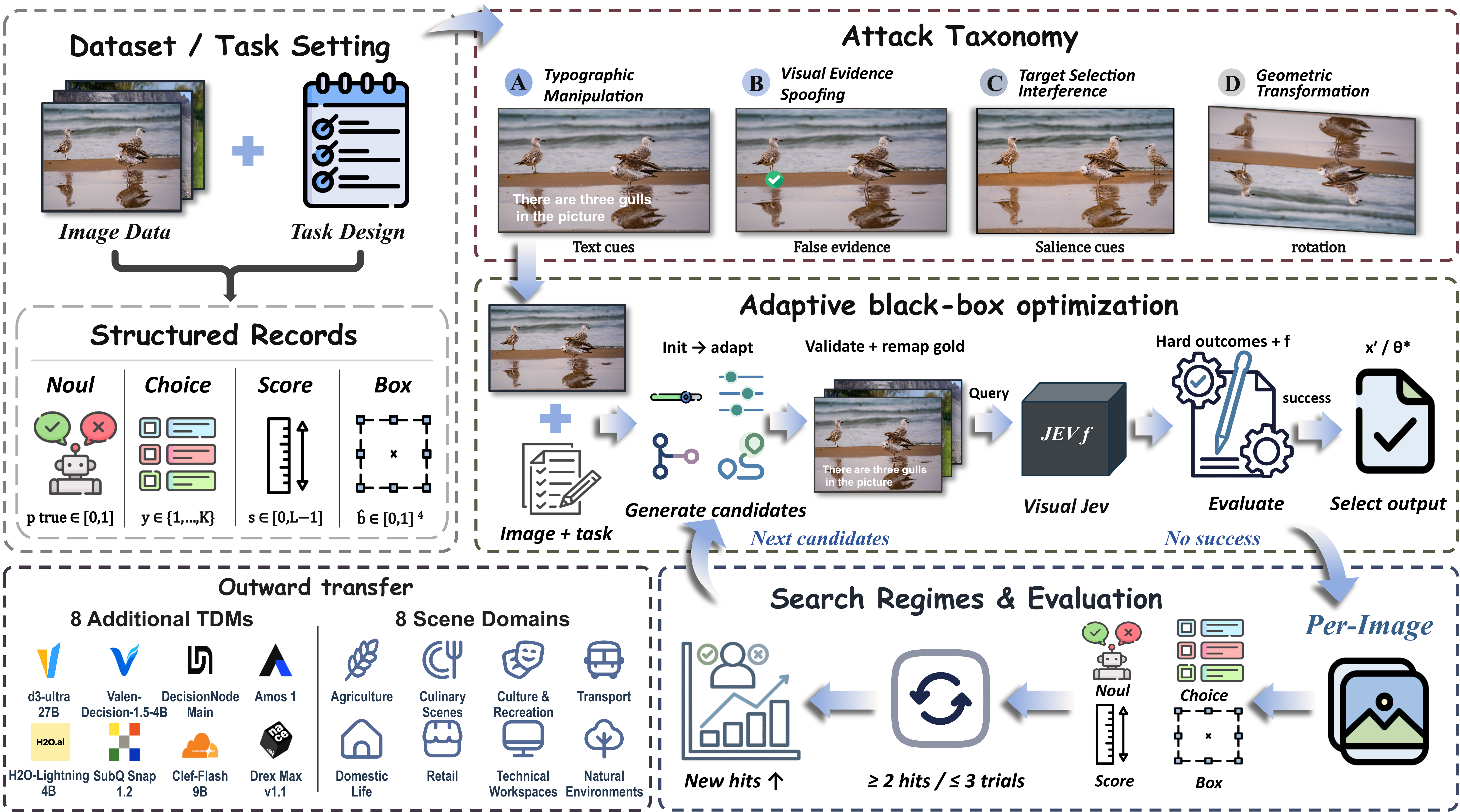}
    }{
        \fbox{%
            \begin{minipage}[c][3.5cm][c]
                {\dimexpr\textwidth-2\fboxsep-2\fboxrule\relax}
                \centering
                \textbf{\ourmethod Pipeline}\\[6pt]
                Parameter proposal
                $\rightarrow$ Candidate rendering
                $\rightarrow$ Jev query
                $\rightarrow$ Feedback and selection\\[6pt]
                Input-specific search
                \qquad / \qquad
                Universal attack search\\[6pt]
                \textit{Pipeline figure placeholder}
            \end{minipage}%
        }
    }
    \caption{
    Overview of Adaptive Black-Box Parameter Search (\ourmethod).
    Jev's responses guide candidate ranking and subsequent
    parameter proposals.
    Input-specific search retains successful attacks for
    individual instances, while universal search selects
    one configuration per attack surface on a training set
    and freezes it for application to unseen images and
    tasks in new scenarios.
    }
    \label{fig:abps_pipeline}
\end{figure*}

\section{Adaptive Black-Box Parameter Search}
\label{sec:method}

The effectiveness of image manipulations depends on their
content, layout, and transformation parameters.
To systematically explore this parameter space, we introduce
\textit{Adaptive Black-Box Parameter Search} (\ourmethod), which
uses typed-output feedback to identify effective image
manipulations.
As illustrated in Fig.~\ref{fig:abps_pipeline}, \ourmethod builds
on promising configurations to generate new candidates
and updates its search history using Jev's responses.
We first describe image-specific search and then extend
the same procedure to universal attacks.

\subsection{Search Space and Feedback}
\label{sec:feedback_search}

\paragraph{Parameterized manipulations.}
A candidate image is generated as $x'=T_m(x;\theta)$,
where $\theta\in\Theta_m$ contains categorical choices and
numerical values discretized into predefined levels.
These parameters control text rendering, evidence-panel
layout, target markings, and geometric transformations.
New candidates build on promising parameter configurations
and are rendered from the original image to avoid
accumulating unintended edits.
\vspace{-5pt}
\paragraph{Typed-output feedback.}
\ourmethod prioritizes attack success and uses numerical
feedback to distinguish candidates with the same success
measure.
For \Noul{}, the feedback is $F_\tau=p_w$, where $p_w$
is the probability of the incorrect binary label.
For \Choice{}, $F_\tau=1-p_y$, where $p_y$ is the
reference-option probability.
For \Score{}, $F_\tau=\min(1,\,|2v-l-h|/(h-l))$,
where $v$ is the returned score and $[l,h]$ is the
acceptable interval with $l<h$.
For \BoxType{}, $F_\tau=1-\operatorname{IoU}(\hat{b},b_\theta)$,
where $\hat{b}$ and $b_\theta$ are the predicted and
reference boxes on the manipulated image.
Higher feedback favors more promising candidates, including
those that have not yet caused an incorrect decision.
Attack success is determined separately using the
task-specific criteria in Sec.~\ref{sec:exp_setup}.

\subsection{Image-Specific Search}
\label{sec:input_search}

For each image condition, \ourmethod starts with predefined valid
configurations and maintains a separate search history.
It evaluates candidate manipulations, uses their results
to select promising parents, and generates new configurations
for questions that remain unresolved.
Algorithm~\ref{alg:abps} summarizes this process.
\vspace{-7pt}
\paragraph{Initialization and candidate generation.}
The initial candidate set is drawn from predefined legal
parameter combinations.
After the first round, new candidates are generated from
high-ranking configurations in the search history.
The proposal function uses four operations:

\begin{tcolorbox}[
    enhanced,
    colback=outputbg,
    colframe=outputborder,
    boxrule=0.6pt,
    arc=5pt,
    left=6pt,
    right=6pt,
    top=5pt,
    bottom=5pt,
    before skip=7pt,
    after skip=7pt,
    fontupper=\normalfont\rmfamily\small,
    boxsep=0pt
]
\noindent
\textbf{Local mutation:}
\textit{Changes one parameter.}\par
\smallskip
\noindent
\textbf{Joint mutation:}
\textit{Changes a predefined parameter pair.}\par
\smallskip
\noindent
\textbf{Recombination:}
\textit{Combines values from two parents.}\par
\smallskip
\noindent
\textbf{Exploration:}
\textit{Samples configurations far from previous trials.}
\end{tcolorbox}

Numerical mutations use neighboring parameter levels,
while categorical mutations select alternative choices.
Recombination inherits values from two parents, and
exploration introduces less-tested combinations.
If an operation cannot produce a valid unseen configuration,
another operation is used.
The resulting candidates are checked for validity before
being submitted to Visual Jev.

\vspace{-7pt}
\paragraph{Evaluation and ranking.}
For each valid candidate, \ourmethod records Jev's response,
attack successes, and typed-output feedback.
Let $\mathcal{R}$ denote the unresolved question types for the
current condition prior to evaluating $\theta$.
Instantiating the attack objective in Eq.~\eqref{eq:attack_objective},
candidates are ranked using
\begin{equation}
    K(\theta) = \big(n(\theta),\, \tfrac{1}{|\mathcal{R}|}\sum\nolimits_{\tau\in\mathcal{R}} F_\tau(\theta)\big),
    \label{eq:instance_ranking}
\end{equation}
where $n(\theta)$ records the number of questions first
attacked when the candidate was evaluated.
The two components are compared lexicographically:
new attack successes take priority, and mean feedback
breaks ties.

\vspace{-7pt}
\paragraph{Search update.}
The evaluation records are added to the search history
$\mathcal{H}^{t}$.
\ourmethod then selects high-ranking parents $\mathcal{P}^{t}$
and proposes the next candidate set:
\begin{equation}
    \mathcal{P}^{t}=\operatorname{Select}(\mathcal{H}^{t};K),
    \qquad
    \mathcal{C}^{t+1}=G(\mathcal{P}^{t},\mathcal{H}^{t}),
    \label{eq:search_update}
\end{equation}
where $G$ applies the four proposal operations within
$\Theta_m$.
The next round repeats rendering and evaluation with these
new parameters.
Earlier evaluations remain available for parent selection,
so each round continues the existing search rather than
restarting it.
\vspace{-1em}
\paragraph{Success confirmation and retention.}
A candidate succeeds when a valid manipulation turns
a correct decision into an incorrect one.
Successful candidates are retained after at least two
incorrect responses in up to three evaluations of
the same image and question. Query budgets, request accounting, and stopping criteria
are detailed in the Appendix.

\begin{algorithm}[t]
\caption{\textit{\ourmethod: Image-Specific Search}}
\label{alg:abps}
\small
\begin{algorithmic}[1]
\Require Image $x$, operator $m$, space $\Theta_m$,
budget $B$, existing successes $\mathcal{S}_0$
\Ensure Retained attacks $\mathcal{S}$

\State $\mathcal{H}\gets\varnothing$;
       $\mathcal{S}\gets\mathcal{S}_0$;
       $\mathcal{C}\gets\operatorname{Init}(\Theta_m)$
\While{$\neg\operatorname{Stop}(\mathcal{H},B)$}
    \ForAll{$\theta\in\mathcal{C}$}
        \State $x'\gets T_m(x;\theta)$
        \Comment{Render candidate}
        \If{$\operatorname{Valid}(x')$ and budget remains}
            \State $e_\theta\gets\operatorname{Evaluate}(x';B)$
            \Comment{Success and feedback}
            \State $\mathcal{H}\gets\mathcal{H}\cup\{(\theta,e_\theta)\}$
        \EndIf
    \EndFor
    \State $\mathcal{S}\gets\mathcal{S}\cup\operatorname{Confirm}(\mathcal{H};B)$
    \Comment{Retain confirmed attacks}
    \If{$\operatorname{Stop}(\mathcal{H},B)$}
        \State \textbf{break}
    \EndIf
    \State $\mathcal{P}\gets\operatorname{Select}(\mathcal{H};K)$
    \State $\mathcal{C}\gets G(\mathcal{P},\mathcal{H})$
    \Comment{Refine and explore}
\EndWhile
\State \Return $\mathcal{S}$
\end{algorithmic}
\end{algorithm}

\subsection{Universal Attack Search}
\label{sec:shared_search}

To extend image-specific search to universal attacks, we
evaluate each candidate configuration on all supported
training images rather than a single image.
Candidate generation remains unchanged, while selection uses the mean attack success rate (ASR) across supported question types:
\begin{equation}
    \theta^\star
    \in
    \underset{\theta\in\mathcal{V}}{\arg\max}
    \frac{1}{|\mathcal{Q}|}
    \sum_{\tau\in\mathcal{Q}}
    \operatorname{ASR}_{\tau}(\theta;\mathcal{D}),
    \label{eq:shared_objective}
\end{equation}
where $\mathcal{D}$ is the training set, $\mathcal{Q}$ contains the
supported question types, and $\mathcal{V}$ contains configurations
with complete, valid evaluations.
Once selected, the winning configuration for each attack surface
is frozen and directly applied to unseen images and new scenarios
without further search. Image-dependent adaptation follows rules fixed during
training, without using target-model feedback to revise
the configuration.
The adaptation rules and information available at
test time are detailed in
Appendix.

\begin{table*}[t]
\centering
\caption{
Attack success rates (\%) across models, attack surfaces,
and question types.
Static Attacks uses fixed manipulation parameters.
\ourmethod retains these successes and adds newly confirmed
attacks found through image-specific search.
Overall averages the applicable attack surfaces for each
question type.
Results of \ourmethod are shown in bold.
A dash indicates unsupported or unavailable results.
}
\label{tab:main_results}

\setlength{\tabcolsep}{3pt}
\renewcommand{\arraystretch}{1.2}

\def\vjbhead#1{%
    \shortstack[c]{%
        \parbox[c][24pt][c]{96pt}{\centering\bfseries #1}\\[-3pt]
        \rule{96pt}{0.35pt}%
    }%
}

\resizebox{\textwidth}{!}{%
\begin{threeparttable}
\begin{tabular}{
    >{\raggedright\arraybackslash}p{105pt}
    l*{20}{c}
}
\toprule

\rowcolor{vjbHeader}
&
& \multicolumn{4}{c}{
    \vjbhead{Typographic\\Manipulation}}
& \multicolumn{4}{c}{
    \vjbhead{Visual Evidence\\Spoofing}}
& \multicolumn{4}{c}{
    \vjbhead{Target Selection\\Interference}}
& \multicolumn{4}{c}{
    \vjbhead{Geometric\\Transformation}}
& \multicolumn{4}{c}{
    \vjbhead{Overall}} \\

\rowcolor{vjbHeader}
\multirow{-2}{*}[4pt]{\textbf{Model}}
& \multirow{-2}{*}[4pt]{\textbf{Method}}
& \Noul{} & \Choice{} & \Score{} & \BoxType{}
& \Noul{} & \Choice{} & \Score{} & \BoxType{}
& \Noul{} & \Choice{} & \Score{} & \BoxType{}
& \Noul{} & \Choice{} & \Score{} & \BoxType{}
& \Noul{} & \Choice{} & \Score{} & \BoxType{} \\
\midrule

& Static Attacks
& 38.80 & 43.20 & 47.90 & 24.30
& 40.70 & 36.00 & 41.80 & 19.00
& 32.50 & 45.10 & 37.60 & 34.30
& 27.40 & 29.30 & 24.50 & 55.10
& 34.90 & 38.40 & 38.00 & 33.20 \\

\rowcolor{vjbPurple}
\cellcolor{white}
\multirow{-2}{105pt}{%
    \raggedright
    \ct{jev-1.14.0}~\cite{typesafe2026jev114}}
& \textbf{\ourmethod (Ours)}
& \textbf{45.90} & \textbf{49.50} & \textbf{59.40} & \textbf{52.90}
& \textbf{46.90} & \textbf{59.60} & \textbf{52.70} & \textbf{31.70}
& \textbf{47.00} & \textbf{58.20} & \textbf{51.60} & \textbf{53.70}
& \textbf{39.30} & \textbf{52.20} & \textbf{41.50} & \textbf{69.60}
& \textbf{44.80} & \textbf{54.90} & \textbf{51.30} & \textbf{52.00} \\
\midrule

& Static Attacks
& 19.20 & 20.70 & 23.80 & 17.60
& 9.60  & 18.60 & 18.70 & 9.40
& 13.50 & 17.00 & 13.00 & 22.90
& 17.50 & 13.50 & 23.10 & 43.70
& 15.00 & 17.50 & 19.70 & 23.40 \\

\rowcolor{vjbPurple}
\cellcolor{white}
\multirow{-2}{105pt}{%
    \raggedright
    \ct{d3-ultra}~\cite{vllmsr2026d3ultra}}
& \textbf{\ourmethod (Ours)}
& \textbf{41.40} & \textbf{34.80} & \textbf{36.20} & \textbf{50.00}
& \textbf{24.50} & \textbf{33.70} & \textbf{33.30} & \textbf{20.30}
& \textbf{15.60} & \textbf{26.10} & \textbf{20.80} & \textbf{32.90}
& \textbf{30.90} & \textbf{30.30} & \textbf{37.20} & \textbf{62.00}
& \textbf{28.10} & \textbf{31.20} & \textbf{31.90} & \textbf{41.30} \\
\midrule

& Static Attacks
& 23.80 & 20.50 & 20.30 & 13.30
& 20.50 & 13.40 & 14.90 & 13.90
& 15.80 & 14.30 & 15.60 & 20.30
& 10.40 & 17.60 & 11.50 & 29.30
& 17.60 & 16.50 & 15.60 & 19.20 \\

\rowcolor{vjbPurple}
\cellcolor{white}
\multirow{-2}{105pt}{%
    \raggedright
    \ct{Valen-Decision-}\newline
    \ct{1.5-4B}~\cite{valen2026model}}
& \textbf{\ourmethod (Ours)}
& \textbf{27.50} & \textbf{28.40} & \textbf{21.50} & \textbf{21.30}
& \textbf{27.40} & \textbf{35.40} & \textbf{23.00} & \textbf{20.80}
& \textbf{26.30} & \textbf{25.00} & \textbf{29.90} & \textbf{28.40}
& \textbf{33.80} & \textbf{37.60} & \textbf{30.80} & \textbf{58.70}
& \textbf{28.80} & \textbf{31.60} & \textbf{26.30} & \textbf{32.30} \\
\midrule

& Static Attacks
& 27.40 & 29.10 & 21.90 & 16.90
& 18.00 & 21.00 & 20.60 & 8.10
& 16.70 & 34.90 & 26.80 & 15.20
& 22.00 & 23.80 & 20.80 & 33.80
& 21.00 & 27.20 & 22.50 & 18.50 \\

\rowcolor{vjbPurple}
\cellcolor{white}
\multirow{-2}{105pt}{%
    \raggedright
    \ct{DecisionNode Main}~\cite{bynn2026decisionnode}}
& \textbf{\ourmethod (Ours)}
& \textbf{34.70} & \textbf{36.00} & \textbf{27.40} & \textbf{27.30}
& \textbf{28.10} & \textbf{42.00} & \textbf{27.90} & \textbf{21.60}
& \textbf{26.70} & \textbf{51.80} & \textbf{43.70} & \textbf{29.10}
& \textbf{41.80} & \textbf{50.00} & \textbf{40.30} & \textbf{56.20}
& \textbf{32.80} & \textbf{45.00} & \textbf{34.80} & \textbf{33.60} \\
\midrule

& Static Attacks
& 38.90 & 32.70 & 44.90 & 60.00
& 19.80 & 27.20 & 16.70 & 78.60
& 16.70 & 12.80 & 14.00 & 35.70
& 16.80 & 17.90 & 21.80 & 78.60
& 23.10 & 22.70 & 24.40 & 63.20 \\

\rowcolor{vjbPurple}
\cellcolor{white}
\multirow{-2}{105pt}{%
    \raggedright
    \ct{Amos 1 Large}~\cite{mukaddam2026amoslarge}}
& \textbf{\ourmethod (Ours)}
& \textbf{61.10} & \textbf{50.00} & \textbf{61.80} & \textbf{80.00}
& \textbf{34.10} & \textbf{48.90} & \textbf{29.80} & \textbf{85.70}
& \textbf{20.00} & \textbf{19.10} & \textbf{20.90} & \textbf{78.60}
& \textbf{31.60} & \textbf{41.10} & \textbf{40.20} & \textbf{100.00}
& \textbf{36.70} & \textbf{39.80} & \textbf{38.20} & \textbf{86.10} \\
\midrule

& Static Attacks
& 32.90 & 26.40 & 31.30 & 25.60
& 32.00 & 29.10 & 25.40 & 13.30
& 27.30 & 26.10 & 24.60 & 11.40
& 14.10 & 19.10 & 16.70 & 33.30
& 26.60 & 25.20 & 24.50 & 20.90 \\

\rowcolor{vjbPurple}
\cellcolor{white}
\multirow{-2}{105pt}{%
    \raggedright
    \ct{H2O-Lightning-}\newline
    \ct{4B}~\cite{h2oai2026lightning}}
& \textbf{\ourmethod (Ours)}
& \textbf{36.60} & \textbf{35.20} & \textbf{34.30} & \textbf{28.20}
& \textbf{41.30} & \textbf{46.50} & \textbf{31.70} & \textbf{33.30}
& \textbf{39.00} & \textbf{38.60} & \textbf{35.40} & \textbf{11.40}
& \textbf{35.90} & \textbf{44.90} & \textbf{34.80} & \textbf{77.80}
& \textbf{38.20} & \textbf{41.30} & \textbf{34.10} & \textbf{37.70} \\
\midrule

& Static Attacks
& 70.80 & 61.20 & 76.90 & -
& 28.60 & 41.90 & 38.70 & -
& 19.40 & 21.10 & 27.30 & -
& 19.10 & 19.80 & 23.10 & -
& 34.50 & 36.00 & 41.50 & - \\

\rowcolor{vjbPurple}
\cellcolor{white}
\multirow{-2}{105pt}{%
    \raggedright
    \ct{SubQ Snap 1.2}%
    \textsuperscript{\(\dagger\)}\hspace{3pt}%
    \cite{subquadratic2026snap12}}
& \textbf{\ourmethod (Ours)}
& \textbf{99.00} & \textbf{80.60} & \textbf{88.50} & -
& \textbf{57.10} & \textbf{51.60} & \textbf{64.00} & -
& \textbf{30.10} & \textbf{33.70} & \textbf{36.40} & -
& \textbf{45.70} & \textbf{47.90} & \textbf{50.00} & -
& \textbf{58.00} & \textbf{53.50} & \textbf{59.70} & - \\
\midrule

%

& Static Attacks
& 41.10 & 23.80 & 48.70 & -
& 40.00 & 25.70 & 50.00 & -
& 20.70 & 13.20 & 44.40 & -
& 23.70 & 19.50 & 43.20 & -
& 31.40 & 20.60 & 46.60 & - \\

\rowcolor{vjbPurple}
\cellcolor{white}
\multirow{-2}{105pt}{%
    \raggedright
    \ct{Clef-Flash 9B}%
    \textsuperscript{\(\dagger\)}\hspace{3pt}%
    \cite{cloudflare2026clefflash}}
& \textbf{\ourmethod (Ours)}
& \textbf{49.50} & \textbf{38.80} & \textbf{51.30} & -
& \textbf{57.80} & \textbf{50.00} & \textbf{64.70} & -
& \textbf{30.40} & \textbf{30.30} & \textbf{47.20} & -
& \textbf{40.90} & \textbf{35.10} & \textbf{54.10} & -
& \textbf{44.60} & \textbf{38.50} & \textbf{54.30} & - \\
\midrule

& Static Attacks
& 53.10 & 54.00 & 59.10 & -
& 23.30 & 37.90 & 33.00 & -
& 14.00 & 15.50 & 13.30 & -
& 18.90 & 12.20 & 15.40 & -
& 27.30 & 29.90 & 30.20 & - \\

\rowcolor{vjbPurple}
\cellcolor{white}
\multirow{-2}{105pt}{%
    \raggedright
    \ct{Drex Max v1.1}%
    \textsuperscript{\(\dagger\)}\hspace{3pt}%
    \cite{nace2026drex}}
& \textbf{\ourmethod (Ours)}
& \textbf{99.00} & \textbf{79.00} & \textbf{77.40} & -
& \textbf{45.60} & \textbf{57.90} & \textbf{52.30} & -
& \textbf{25.80} & \textbf{33.00} & \textbf{22.20} & -
& \textbf{44.20} & \textbf{41.80} & \textbf{44.00} & -
& \textbf{53.70} & \textbf{52.90} & \textbf{49.00} & - \\
\bottomrule

\end{tabular}
\end{threeparttable}%
}

\vspace{3pt}
\begin{minipage}{\textwidth}
\footnotesize
\textsuperscript{\(\dagger\)}
These models are evaluated on \Noul{}, \Choice{}, and
\Score{} only; \BoxType{} evaluation is not applicable.
\end{minipage}

\end{table*}

\section{Experiments}
\label{sec:experiments}

We evaluate the effectiveness of direct image manipulations
against Visual Jev and examine their generalization to
unseen images and new scenarios.
Our experiments address three research questions:

\begin{tcolorbox}[
    enhanced,
    colback=rqbodybg,
    colframe=rqborder,
    colbacktitle=rqtitlebg,
    coltitle=rqborder,
    title={Key Research Questions},
    fonttitle=\bfseries\itshape,
    fontupper=\normalfont\rmfamily,
    boxrule=0.7pt,
    arc=2pt,
    titlerule=0.6pt,
    left=6pt,
    right=6pt,
    top=5pt,
    bottom=5pt,
    toptitle=4pt,
    bottomtitle=4pt,
    boxsep=0pt,
    before skip=7pt,
    after skip=8pt
]
\noindent
\textbf{RQ1:}
\textit{How vulnerable are Visual Jev's typed decisions
to direct image manipulations?}\par
\smallskip
\noindent
\textbf{RQ2:}
\textit{Can a fixed manipulation induce incorrect decisions
on unseen images?}\par
\smallskip
\noindent
\textbf{RQ3:}
\textit{Do manipulations found in one scenario
transfer to other scenarios?}
\end{tcolorbox}

\subsection{Experimental Setup}
\label{sec:exp_setup}

\paragraph{Datasets.}
We evaluate on \textsc{VisualJevBench}, which contains
800 images and 3,200 questions across eight scenarios
and covers \Noul{}, \Choice{}, \Score{}, and \BoxType{}.
Training and held-out sets are separated by source image,
with all questions and manipulated variants of an image
assigned to the same split.
\vspace{-9pt}
{\setlength{\emergencystretch}{2em}

\paragraph{Target models.}
Our primary target is \ct{jev-1.14.0}, accessed through the official TypeSafe API~\cite{typesafe_docs}. We additionally evaluate \ct{d3-ultra 27B}~\cite{vllmsr2026d3ultra}, \ct{Valen-Decision-\allowbreak 1.5-4B}~\cite{valen2026model}, \ct{DecisionNode Main}~\cite{bynn2026decisionnode}, \ct{Amos 1 Large}~\cite{mukaddam2026amoslarge}, \ct{H2O-Lightning-\allowbreak 4B}~\cite{h2oai2026lightning}, \ct{SubQ Snap 1.2}~\cite{subquadratic2026snap12}, \ct{Clef-Flash 9B}~\cite{cloudflare2026clefflash}, and \ct{Drex Max v1.1}~\cite{nace2026drex}. Each model's version and inference settings remain fixed throughout evaluation (please see Appendix).
}

\vspace{-9pt}
\paragraph{Evaluation metrics.}
Following JevAdvBench~\cite{hu2026jevadvbench}, we report
attack success rate (ASR), defined as the proportion of
clean-correct instances for which a valid image manipulation
induces an incorrect decision.
\Noul{} uses its specified probability threshold,
\Choice{} requires agreement with the reference option,
\Score{} uses an acceptable interval or reference-level
criterion, and \BoxType{} uses an IoU threshold against
the reference box on the submitted canvas.
For each question type, Overall averages ASR across
applicable attack surfaces.
Aggregates across question types weight supported types equally.
The benign-input flip rate remains below $2\%$ for
\Noul{}, \Choice{}, and \Score{}, and below $5\%$
for \BoxType{}.
Detailed data splits, search configurations, correctness
thresholds, eligible sample counts, and clean-baseline
results are provided in Appendix.

\subsection{RQ1: Effectiveness of Image Manipulations}
\label{sec:exp_surfaces}

Table~\ref{tab:main_results} compares the unoptimized
\textsc{VisualJevBench} baseline (Static Attacks) with image-specific \ourmethod.

\paragraph{Consistent gains across all question types.}
\ourmethod substantially improves ASR over Static Attacks across all
four question types.
On Visual Jev, Overall ASR rises to $54.90\%$ on \Choice{},
$52.00\%$ on \BoxType{}, $51.30\%$ on \Score{}, and $44.80\%$
on \Noul{}, with the largest gain on \BoxType{} ($+18.80$
percentage points).
Different question types are most vulnerable to different
mechanisms: geometric transformation achieves the highest ASR on
\BoxType{} ($69.60\%$) because rotation and perspective shifts
directly distort spatial cues required for localization, whereas
typographic manipulation ($59.40\%$ on \Score{}) and visual
evidence spoofing ($59.60\%$ on \Choice{}) inject competing
textual or visual evidence that misleads scoring and option
selection.

\paragraph{Consistent effectiveness across models.}
The same advantage holds across all other evaluated models,
demonstrating effectiveness across models.
For instance, \ourmethod lifts Overall \BoxType{} ASR on d3-ultra 27B
from $23.40\%$ to $41.30\%$, while reaching up
to $99.00\%$ ASR on SubQ Snap 1.2 (\Noul{}) under typographic
manipulation.
Overall, black-box parameter search consistently exposes
vulnerabilities beyond fixed manipulations across diverse typed
decision models.

\begin{table}[t]
\centering
\caption{
ASR (\%) of universal attacks on unseen images.
Configurations are frozen after training-set search
and applied without further optimization.
}
\label{tab:universal_results}

\small
\setlength{\tabcolsep}{4pt}
\renewcommand{\arraystretch}{1.3}

\resizebox{\columnwidth}{!}{%
\begin{tabular}{lcccc}
\toprule
\rowcolor{vjbHeader}
\textbf{Attack surface}
& \textbf{\Noul{}}
& \textbf{\Choice{}}
& \textbf{\Score{}}
& \textbf{\BoxType{}} \\
\midrule

Typographic Manipulation
& 16.90 & 27.20 & 16.30 & 22.10 \\

Visual Evidence Spoofing
& 24.60 & 27.00 & 21.00 & 38.70 \\

Target Selection Interference
& 32.20 & 73.00 & 33.40 & 34.40 \\

Geometric Transformation
& 27.60 & 36.80 & 28.80 & 70.40 \\

\bottomrule
\end{tabular}%
}
\end{table}

\subsection{RQ2: Generalization to Unseen Images}
\label{sec:exp_universal}

We next evaluate whether a shared manipulation frozen from the
training set remains effective on unseen images within the same
scenarios without further optimization (Table~\ref{tab:universal_results}).
Averaged across all four attack surfaces and four question types,
universal attacks achieve $33.15\%$ ASR, confirming that fixed
manipulations transfer to new images without per-image search.
Among the four surfaces, target selection interference ($43.25\%$)
and geometric transformation ($40.90\%$) generalize substantially
better than visual evidence spoofing ($27.83\%$) and typographic
manipulation ($20.63\%$), as distractor highlights and spatial
distortions rely less on image-specific semantics than fabricated
evidence or injected text.

Moreover, the strongest universal attack depends on the target
question type.
Target selection interference reaches $73.00\%$ ASR on \Choice{}
(compared to $34.40\%$ on \BoxType{}), whereas geometric
transformation reaches $70.40\%$ on \BoxType{} (compared to
$36.80\%$ on \Choice{}).
Thus, even without per-image tuning, reusable manipulations
remain highly effective when their mechanism aligns with the
structured decision interface being evaluated.

\begin{figure}[t]
    \centering
    \includegraphics[width=\columnwidth]
        {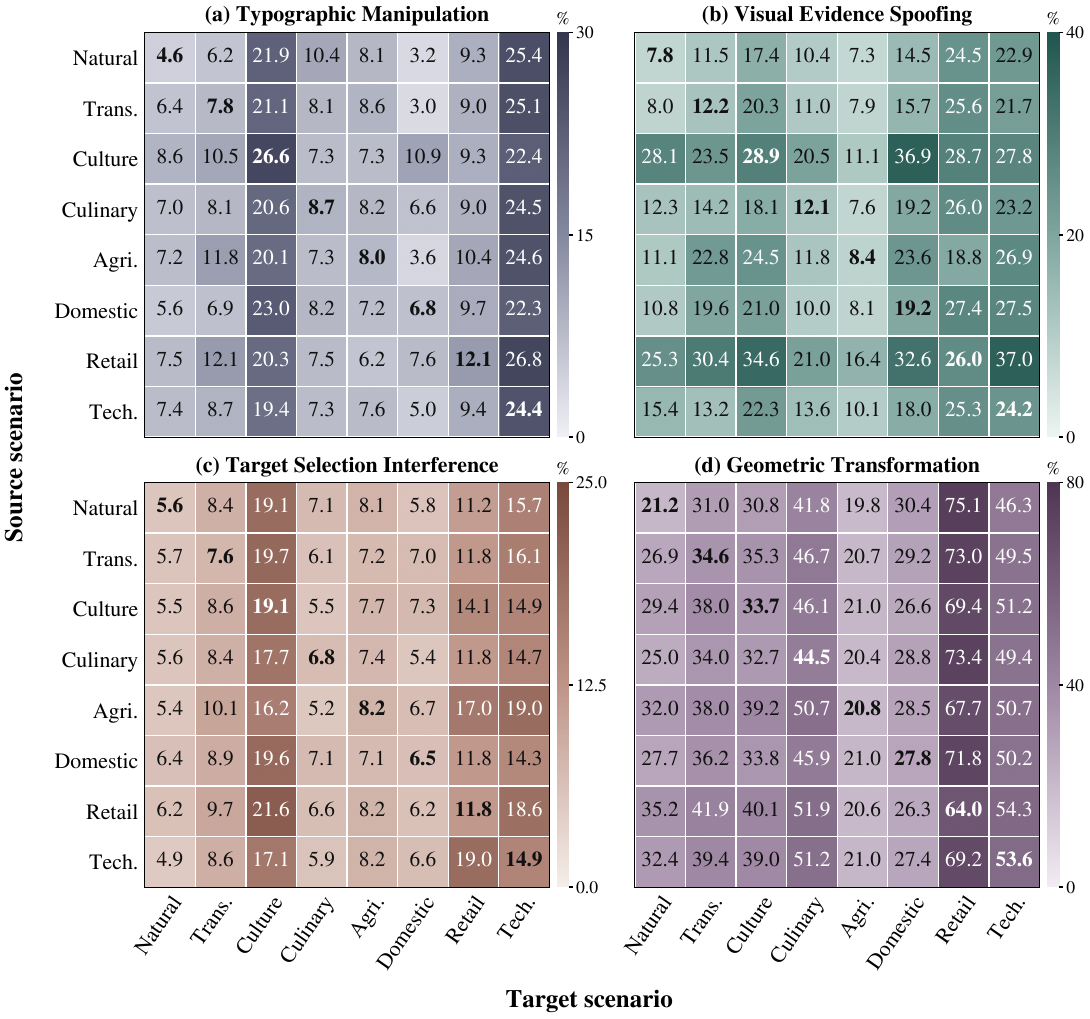}
    \caption{
    Cross-scenario ASR (\%) across four attack surfaces
    with frozen parameters and adaptation rules.
    Rows/columns indicate source/target scenarios;
    diagonal entries are bold.
    Panels use separate color scales; darker indicates higher ASR.
    Scenario names are shortened to their first word,
with abbreviations where needed.
    }
    \label{fig:task_transfer}
    \vspace{-5pt}
\end{figure}

\subsection{RQ3: Transfer Across Scenarios}
\label{sec:exp_transfer}

To answer RQ3, we apply the frozen configuration from
each source scenario to other target scenarios.
As shown in Fig.~\ref{fig:task_transfer}, geometric
transformation achieves the highest unweighted
off-diagonal mean ASR of $40.09\%$ across attack surfaces.
This suggests that changes in spatial presentation
transfer more readily across scenarios than
content-dependent distractor markings or injected text.

\begin{table}[t]
\centering
\caption{
Ablation results on Visual Jev under the same query budget.
Avg. is mean ASR (\%) across question types;
Queries is the mean API query count per image.
}
\label{tab:ablation}
\footnotesize
\setlength{\tabcolsep}{3pt}
\resizebox{\linewidth}{!}{%
\begin{tabular}{lcccccc}
\toprule
Variant & \Noul{} & \Choice{} & \Score{} & \BoxType{}
& Avg. ($\uparrow$) & Queries ($\downarrow$) \\
\midrule
w/o Feedback
& 36.40 & 41.20 & 40.10 & 36.80 & 38.63 & 13.6 \\
w/o Joint \& Recomb.
& 38.90 & 45.60 & 43.80 & 42.50 & 42.70 & 11.4 \\
w/o Exploration
& 40.80 & 48.50 & 46.20 & 45.70 & 45.30 & 10.5 \\
\midrule
\rowcolor{vjbPurple}
\textbf{\ourmethod (Ours)}
& \textbf{44.80} & \textbf{54.90}
& \textbf{51.30} & \textbf{52.00}
& \textbf{50.75} & \textbf{7.2} \\
\bottomrule
\end{tabular}%
}
\vspace{-10pt}
\end{table}

Transferability also varies substantially with the
target scenario.
For geometric transformation, off-diagonal ASR ranges
from $67.70\%$ to $75.10\%$ on Retail but only
$19.80\%$ to $21.00\%$ on Agriculture across source
scenarios.
This variation suggests that cross-scenario success
depends on the target domain's visual characteristics.
Overall, these results show that the vulnerabilities
exposed by \textsc{VisualJevBench} extend beyond
individual scenarios, with fixed manipulations
transferring across distinct visual domains.

\subsection{Defense Evaluation}
\label{sec:exp_defense}

We evaluate six defenses on valid successful attacks:
CLIP-based filtering~\cite{compvis2022safetychecker},
Q16~\cite{schramowski2022q16},
JPEG compression~\cite{guo2018countering},
median filtering~\cite{xu2018feature},
resizing~\cite{gushchin2025guardians},
and color quantization~\cite{xu2018feature}.
We measure interception rates on successful attacks
and false positive rates on clean-correct inputs.

\begin{figure}[t]
\centering
\includegraphics[width=\linewidth]{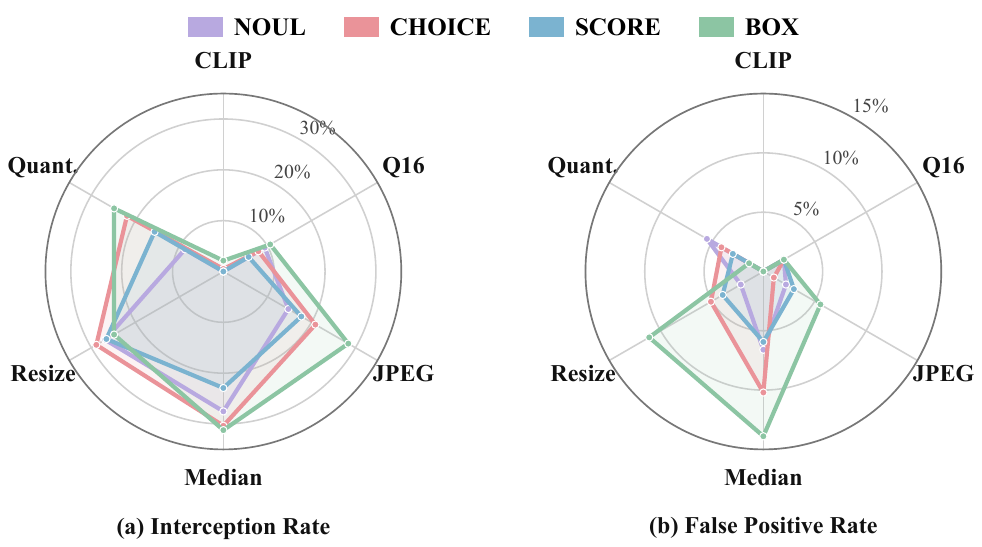}
\caption{
Six defenses on Visual Jev:
(a) interception rate on successful attacks;
(b) false positive rate on
clean-correct inputs.
}
\label{fig:defense_radar}
\vspace{-10pt}
\end{figure}

Figure~\ref{fig:defense_radar} shows that safety
filters intercept fewer than $9\%$ of attacks on average.
Preprocessing achieves $17.77$--$28.00\%$ interception,
but can impair clean performance.
Median filtering achieves the highest average
interception rate, while turning $13.89\%$ of initially
correct \BoxType{} decisions into incorrect ones.

\subsection{Ablation Study and Query Efficiency}
\label{sec:exp_ablation}

To quantify the contribution of each search component in \ourmethod,
we conduct an ablation study on Visual Jev under the same query
budget $B$ (Table~\ref{tab:ablation}), individually removing
typed-output feedback ($F_\tau$, Sec.~\ref{sec:feedback_search}),
joint mutation and recombination (Sec.~\ref{sec:input_search}),
and exploration (Sec.~\ref{sec:input_search}).
Removing typed-output feedback causes the steepest drop, reducing average ASR by $12.12$ percentage points, confirming that continuous probability, score, and IoU signals are essential for ranking promising candidates before an attack succeeds.
Disabling joint mutation and recombination lowers average ASR to $42.70\%$ by missing coupled parameter interactions, while removing exploration drops it to $45.30\%$ as the search becomes trapped in local optima.

Beyond higher attack effectiveness, \ourmethod also substantially improves query efficiency.
Full \ourmethod requires only $7.2$ API queries per image
on average, compared with $10.5$--$13.6$ for the ablated
variants, highlighting the query efficiency of its
combined search components. We additionally compare \ourmethod with Random Search
under matched query budgets and evaluation conditions
(please see Appendix C).

\section{Conclusion, Limitations, and Future Work}

We introduce \textsc{VisualJevBench} and \ourmethod
to evaluate image-only attacks on
vision-conditioned typed decision models.
Experiments reveal vulnerabilities to
direct image manipulations, with attacks generalizing
to unseen images and transferring across scenarios.
Existing image-level defenses offer only partial protection.
Our evaluation is limited to predefined manipulation
families and non-adaptive defense testing.
Future work will explore broader attack strategies
and defenses against adaptive attacks.

\section*{LLM Usage Statement}

GPT-5.6 was used solely to assist with manuscript language editing (improving readability and clarity) and minor code debugging and refactoring. All benchmark images and annotations in \textsc{VisualJevBench} were sourced from existing datasets and manually curated and verified by the authors. AI tools were not used to propose research hypotheses, design the benchmark or the ABPS framework, conduct experiments, or interpret experimental results.

All AI-assisted text edits and code modifications were manually inspected and validated by the authors using the same experimental pipeline applied to the rest of the implementation. The authors take full responsibility for the accuracy, integrity, and reproducibility of the final work, including the dataset, code, experimental results, and conclusions.
{
    \small
    \bibliographystyle{ieeenat_fullname}
    \bibliography{main}
}

\section{Additional Related Work}
\label{app:related_work}

\paragraph{Adversarial robustness of visual models.}
Prior work has investigated the robustness and security of visual models under a variety of adversarial settings. Existing studies cover defenses against malicious image manipulation~\cite{zeng2025psfd}, adversarial vulnerabilities and backdoor detection in vision-based systems~\cite{zhang2024detector,zhang2025test,wang2024trojanrobot}, and security challenges involving learned visual representations~\cite{wang2026dual,wang2024unlearnable,zhou2025darkhash,yao2024reverse}. Related research has also examined adversarial attacks and defense mechanisms in language-based and task-oriented AI systems~\cite{zhang2026defending,wang2026pvdetector}. Although these studies consider different threat models and attack surfaces, they collectively highlight the importance of evaluating model robustness under adversarial conditions. Our work focuses specifically on how direct image manipulations affect structured visual decisions under black-box access.

\end{document}